\documentclass[useAMS,a4paper,usenatbib]{mnras}

\usepackage{mathptmx}

\usepackage[T1]{fontenc}
\usepackage{ae,aecompl}

\usepackage{graphicx}
\usepackage{amsmath}
\usepackage{amssymb}
\usepackage{color}

\usepackage{fixltx2e}
\usepackage{hyperref}
\usepackage{breakurl}

\newcommand{\kom}[1]{#1}

\title{Observation of the Black Widow B1957+20 millisecond pulsar binary system with the MAGIC telescopes}

\author[M.~L.~Ahnen~et.~al.]
{M.~L.~Ahnen$^{1}$, 
S.~Ansoldi$^{2,25}$, 
L.~A.~Antonelli$^{3}$, 
C.~Arcaro$^{4}$, 
A.~Babi\'c$^{5}$, 
B.~Banerjee$^{6}$,\newauthor 
P.~Bangale$^{7}$, 
U.~Barres de Almeida$^{7,26}$
J.~A.~Barrio$^{8}$
% ...
(Affiliations can be found after the references)
}

\begin{document} 

\date{Accepted . Received ; in original form }
\pagerange{\pageref{firstpage}--\pageref{lastpage}} \pubyear{2017}

\maketitle

\label{firstpage}

  \begin{abstract}
B1957+20 is a millisecond pulsar located in a black widow type compact binary system with a low mass stellar companion. 
The interaction of the pulsar wind with the companion star wind and/or the interstellar plasma is expected to create plausible conditions for acceleration of electrons to TeV energies and subsequent production of very high energy $\gamma$ rays in the inverse Compton process. 
We performed extensive observations with the MAGIC telescopes of B1957+20. 
We interpret results in the framework of a few different models, namely emission from the vicinity of the millisecond pulsar, the interaction of the pulsar and stellar companion wind region, or bow shock nebula.
No significant steady very high energy $\gamma$-ray emission was found.
We derived a 95\% confidence level upper limit of $3.0 \times 10^{-12}\,\mathrm{cm^{-2}\,s^{-1}}$ on the average $\gamma$-ray emission from the binary system above 200 GeV. 
The upper limits obtained with MAGIC constrain, for the first time, different models of the high-energy emission in B1957+20. 
In particular, in the inner mixed wind nebula model with mono-energetic injection of electrons, the acceleration efficiency of electrons is constrained to be below $\sim(2-10)$\% of the pulsar spin down power.
For the pulsar emission, the obtained upper limits for each emission peak are well above the exponential cut-off fits to the {\it Fermi}-LAT data, extrapolated to energies above 50 GeV. 
The MAGIC upper limits can rule out a simple power-law tail extension through the sub-TeV energy range for the main peak seen at radio frequencies.
\end{abstract}


\section{Introduction}

B1957+20 is a millisecond pulsar (MSP), with a period of 1.6\,ms and a surface magnetic field $\sim$10$^8$ G. 
It is located within a very compact binary system of the black widow type with a period of 9.2 h (\citealp{fru88}).  
The black widow binary systems contain a very low mass companion star which is being evaporated due to the energy release by the MSP. 
In the case of B1957+20 the companion star has the mass of $\sim\,0.022$ M$_\odot$~(\citealp{van88}) and a radius of $\sim10^{10}$ cm.
Its surface temperature varies between 2900 K and 8300 K for the side illuminated by the pulsar \citep{fru96, rey07}. The pulsed radio emission of B1957+20 is eclipsed for about $10\%$ of the orbital phase by the wind from the companion star. The wind region is limited by the shock formed in the interaction of the MSP and companion winds within the binary system \citep{fru88}. 
The existence of the wind collision shock is also supported by the observations of the X-ray emission, modulated with the period of the binary system \citep{hua12}. The model for the X-ray emission in the wind shock scenario has been considered by \citet{hg90} and \citet{at93} and more recently by \citet{whvb15}. 

The binary system B1957+20 is surrounded by an H$\alpha$ emission nebula which is expected to be produced in the interaction of the pulsar wind with the interstellar medium \citep{kh88}.  
The nebula reveals a bow shock around the system, which moves through the medium with a velocity of $\sim220$ km s$^{-1}$, based on the proper motion measurements \citep{arz94}. The extended X-ray emission has \kom{also been} reported from the interior of the bow shock \citep{sta03, hb07, hua12}. 
This non-thermal X-ray emission creates a tail behind the moving pulsar with a length of $\sim\,10^{18}$ cm. 
The extended emission is interpreted as produced by relativistic electrons accelerated in the pulsar vicinity \citep{che06}.

This binary system was claimed in the past to be a GeV-TeV $\gamma$-ray source by the Potchefstroom air Cerenkov telescope and the \textit{COS B} satellite \citep{bri90}. 
These findings have not been confirmed by the Potchefstroom/Nooitgedacht \citep{rau95} and the EGRET observations \citep{buc96}. 
Such potential, non-pulsed emission might be produced by the comptonization of the companion star radiation by leptons accelerated in the collision region between the pulsar wind and the donor star wind \citep{rau95}.
A pulsed GeV $\gamma$-ray emission was discovered by the {\it Fermi}-LAT \citep{gu12}. This emission shows two characteristic peaks in the light curve. 
The spectrum is well described by a power-law with an exponential cut-off at the energy of a few GeV.
Similar cut-offs have also been observed in classical pulsars. 
Therefore, the particle acceleration processes within inner pulsar magnetospheres can be similar for classical and millisecond pulsars. These similarities may also indicate that the properties of the winds produced by the millisecond and classical pulsars are likely similar. In fact, the classical pulsar PSR B1259-63, 
found within the binary system on the orbit around a Be type star, has been reported to emit TeV $\gamma$ rays close to the periastron passage \citep{aha05}. 
The discovery of TeV emission from a binary system containing millisecond pulsars will put a new constraint on the acceleration and radiation processes of energetic particles in the vicinity of pulsars. 
Due to the different compactness of the TeV $\gamma$-ray and millisecond pulsar type of binaries, as well as the different radiation fields produced by the companion stars, the detection of TeV $\gamma$ rays could provide 
new insight into the structure of the pulsar winds, their magnetization and localization of acceleration regions outside the light cylinder radius.

The existence of an additional higher energy GeV $\gamma$-ray component, which appears at a specific range of the orbital phases, has been also reported based on the analysis of {\it Fermi}-LAT data \citep{wu12}. 
This component can be modelled by electrons from the pulsar comptonizing the radiation from the companion star.

\begin{figure}
  \includegraphics[width=0.49\textwidth]{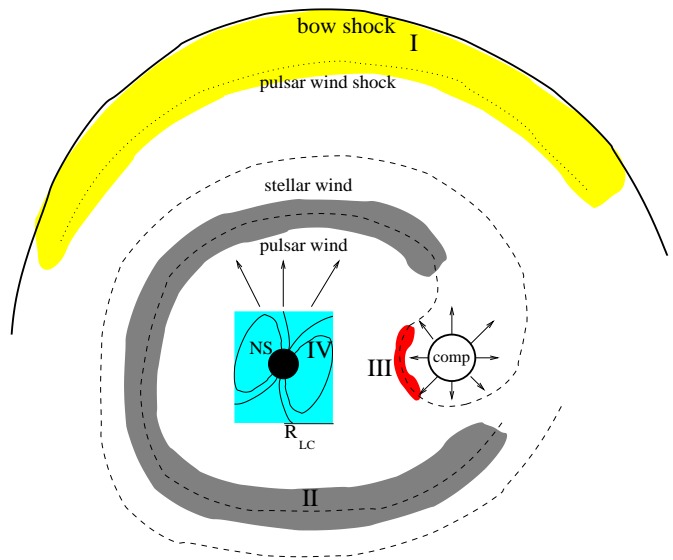}
  \caption{
    Schematic cartoon showing zones of various scenarios of $\gamma$-ray emission from the B1957+20 system. 
    The emission associated with the pulsar wind shock within the bow shock of the binary occurs in region I (yellow, see Sect.~\ref{sec:bowshock}).
    In the region II (in gray) mixing of the pulsar wind with the wind of its stellar companion occurs outside of the binary system (see Sect.~\ref{sec:innernebula}).
    In the region III a modulated emission can be produced from the direct interaction, within the binary system, of the two winds  (red, see Sect.~\ref{sec:innerbinary}).
    The emission might be also generated in the vicinity of the pulsar (cyan, IV, see Sect.~\ref{sec:pulsar}).
  }
  \label{fig:models}
\end{figure}

The variety of the non-thermal phenomena observed in the MSP binary system B1957+20 suggests that the conditions in its vicinity are suitable for production of very high energy (E$>100$\,GeV) $\gamma$ rays.  
In this paper we present the results of the very high energy range $\gamma$-ray observations of B1957+20 performed with the MAGIC telescopes. 
The derived upper limits on the high energy $\gamma$-ray emission from the nebula around the binary system, from the binary system itself and from the millisecond pulsar B1957+20, are discussed in the context of available theoretical models.
In particular, we consider a possible production of the high energy $\gamma$ rays by electrons 
accelerated on the large scale MSP shock in the pulsar wind within the bow shock nebula; 
on the shocks resulted from the mixture of the pulsar and companion star winds in the equatorial region of the binary system (they are twisted around the binary system due to the Coriolis force);
on the shock in the inner part of the binary system formed between the companion star and the pulsar, 
and also $\gamma$ rays which might be produced in the region close to the MSP light cylinder or in the MSP inner magnetosphere.
The different emission zones occurring in those models are schematically depicted in Fig.~\ref{fig:models}.

\section{MAGIC data and analysis}
MAGIC is a system of two Imaging Atmospheric Cherenkov Telescopes located at Roque de los Muchachos, on the Canary island of La Palma, Spain \citep{ale16a}. 
It is used for observations of $\gamma$ rays with energies above $\sim\,$50\,GeV.
It has a sensitivity of $(0.66\pm0.03)$\% of C.U. (Crab Nebula flux) in 50 h of observations for energies above 220\,GeV \citep{ale16b}. 
The angular resolution (defined as a standard deviation of a two-dimensional Gaussian distribution) at those energies is $\lesssim0.07^\circ$.

B1957+20 was observed by the MAGIC telescopes between June and October 2013.
The data were taken at low and \kom{medium} zenith angles, $8^\circ-52^\circ$, with
most of the data taken below $30^\circ$.
We analyzed the data using the standard MAGIC analysis chain \citep{za13, ale16b}.
To the data affected by non-perfect atmospheric conditions we have applied a
correction based on simultaneous LIDAR measurements \citep{fg15}.

Together with each event image, we recorded the absolute event arrival time 
using a GPS receiver. 
The absolute time precision of the events registered by MAGIC is 200\,ns. 

For the analysis of the binary system we have selected 66.5\,h of data based on the rate of background events and atmospheric transmission from a height of 9\,km being above $0.6$ with respect to the perfect atmospheric conditions.
We performed analysis in bins of orbital period using ephemeris of \citet{gu12}. 
The orbital phase is defined starting from the ascending node with respect to the millisecond pulsar, i.e. phase 0.75 corresponds to the pulsar being between companion and the observer.
For the pulsar analysis, however, tighter cuts were applied to assure
the lowest possible energy threshold. Data taken
at zenith angles above 30$^\circ$ were discarded as well as those
with too low atmospheric transmission for pulsar studies (the integrated transmission above 9 km exceeding 0.8 of the one with perfect atmospheric conditions). 
The observation time for the pulsar analysis is 29.4\,h.

The pulsar rotational phase of the events was computed using \textit{Tempo2} \citep{tempo2}, and an ephemeris for PSR~B1957+20 covering the MAGIC observations. 
The ephemeris was constructed by analyzing pulsar timing data recorded at the Nan\c{c}ay radio telescope, near Orl\'{e}ans, France. 
The radio data were scrunched in frequency and in polarization, and were integrated in time in order to build one pulsar profile every 10 minutes of observation. 
A total of 234 times of arrival were then extracted from these data and analyzed with \textit{Tempo2} to obtain a timing solution for PSR~B1957+20 covering the MJD period 56048 to 57270. 
The timing solution uses the BTX binary model and three orbital frequency derivatives, to track changes in the orbital period of the system accurately. 
The weighted rms of the timing residuals is 2.5 $\mu$s. 
Note that this ephemeris has a different phase 0 definition than that used in \cite{gu12}.

We compute the significance using equation 17 in \cite{LiMa}.
All upper limits on the flux presented in this paper were calculated following the approach of \cite{ro05} using 95\% confidence level and assuming 30\% total systematic uncertainty.
For calculation of upper limits on the flux from the binary system we require in addition that the upper limit on the number of excess events is above 3\% of the residual background. 
This ensures that the background induced systematic uncertainties do not exceed the claimed above total systematic uncertainty (see \citealp{ale16b}).
Such additional condition is not needed in the case of pulsar analysis, as the off-pulse background estimation is expected to have a negligible effect on total systematic uncertainty.

\section{\textit{Fermi}-LAT data and analysis}\label{sec:fermiana}
A data sample of more than 3 years (from 56048 up to 57270 MJD) of {\it Fermi}-LAT data was analyzed, according to the period of validity of the ephemeris used.
We analyzed this data set using the  P8R2\_SOURCE\_V6 instrument response
functions and the {\it Fermi} Science Tools version v10r0p5\footnote{Available online at \burl{http://fermi.gsfc.nasa.gov/ssc/data/analysis/software/}}.
Events were selected within a circular region of interest (ROI) of $15^{\circ}$  centered at the
pulsar position (R.A.=$19^h 59^m 36.77^s$ , Dec=$20^\circ 48^{'}
15.12^{''}$). 
We selected ``Source'' class events (evclass = 128 and evtype = 3) that were
recorded only when the telescope was in nominal science mode. 
To reject the background coming from the Earth's limb, we selected photons with a zenith angle $ \leq 90^{\circ}$.
The pulsar rotational phase and barycentric corrections of the events were computed using {\it
  Tempo2} with the {\it Fermi} plug-in\footnote{\burl{http://fermi.gsfc.nasa.gov/ssc/data/analysis/user/Fermi\_plug\_doc.pdf}}
, using the same ephemeris as for MAGIC data analysis.
The pulsar light curve was produced applying an additional energy dependent
angular cut defined by:
 $R = \max(0.8\,(E/1 \mathrm{GeV})^{-0.8},0.2)^{\circ}$, according to the
 approximation of the \textit{Fermi}-LAT
 Pass8 point spread function for a 68\% confinement radius \citep{3FGL}.

For the spectral analysis, a binned likelihood analysis was performed making use of 
the {\it pyLikelihood} python module of the {\it Fermi} tools, 
for both peaks, P1 and P2, separately.
We started by including all sources in ROI from the third {\it Fermi} Source Catalog
\citep{3FGL} in the spectral-spatial model, resulting in a total of 174
 sources. All sources from 3FGL catalog were assumed to have spectral type as
suggested in the catalog. 
The spectral parameters for sources with a significance higher than $5\sigma$ and located at less
than 5 degrees away from the ROI center were left free. 
We also let the normalization factor of the isotropic
(iso\_P8R2\_SOURCE\_V6\_v06.txt) and Galactic (gll\_iem\_v06.fits) background
models free. For the rest of the sources all parameters were left fixed to their catalog value.
Finally, all sources with TS<4 were removed from the model, reducing the
number of sources to 66 and the number of free parameters to 13. 
For the calculation of the spectral points, we repeated the procedure in each energy bin 
using a power-law with the normalization factor free and a spectral index
fixed to $2$. Whenever the significance of the spectral point was less than $1.5\sigma$, a 95\% confidence level upper limit was calculated instead.

In order to search for a possible nebula emission at GeV energies we performed a spectral fit to the photons falling in the off-pulse interval (defined in Table \ref{peakRegions}), where no pulsar emission is expected. For this fit we used a model in which the pulsar component is removed and the nebula is described by a power-law point-like source.

\begin{table}
 \centering 
 \caption{\small Definition of the signal and background regions derived from the \textit{Fermi}-LAT data.}
\begin{tabular}{ccc}
 \hline 
P1& P2 & Off-region\\
\hline
-0.016 - 0.008  &  0.520 - 0.563  &   0.64\,-\,0.94\\
 \hline
 \end{tabular}
\label{peakRegions}
 \end{table}

%

\section{Results}\label{sec:res}
We have performed an analysis of the MAGIC data searching for a steady emission from the Black Widow system.
The pulsed emission is investigated using MAGIC and \textit{Fermi}-LAT data.

\subsection{Analysis of the nebular emission}
\begin{figure}
  \includegraphics[width=0.49\textwidth]{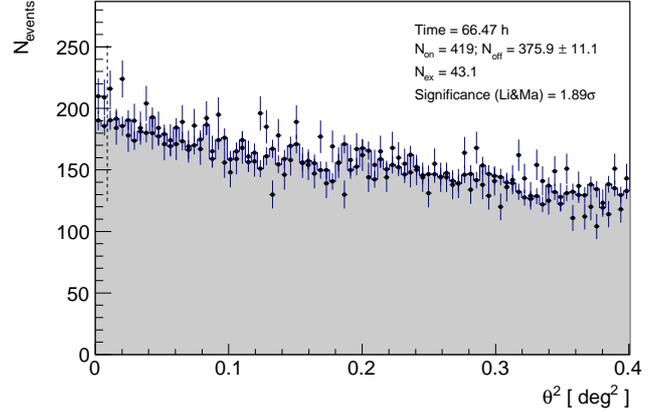}
\caption{
Distribution of the squared angular distance between the \kom{reconstructed event direction and the nominal source position} (points) and the background estimation (shaded area).
The corresponding energy threshold (defined as the peak of the differential energy distribution for Monte Carlo $\gamma$ rays with a spectral slope of $-2.6$) is $\sim\,250\,$GeV.
}\label{fig:th2}
\end{figure}
\begin{figure}
  \includegraphics[width=0.49\textwidth]{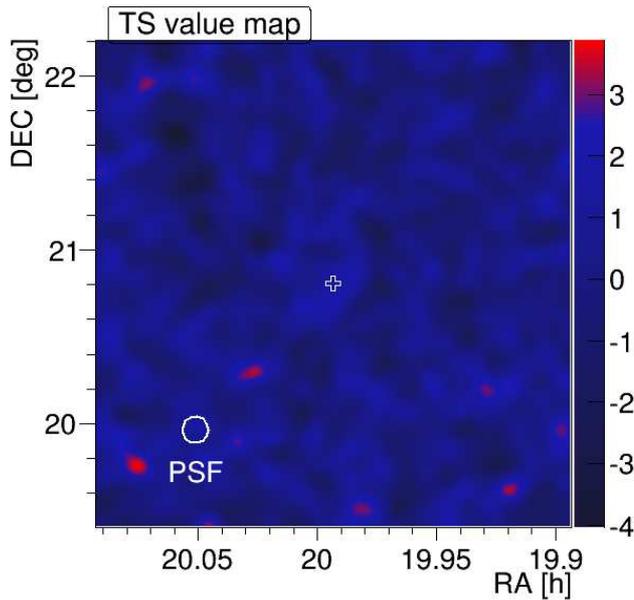}
\caption{
Significance skymap of the region around the position of B1957+20 (marked with an empty cross).
The corresponding energy threshold is $\sim\,250\,$GeV.
}\label{fig:skymap}
\end{figure}

Using the standard detection cuts of MAGIC\footnote{They have been selected such as to provide optimal sensitivity over a broad range of possible spectral shapes of the source. 
}, no significant $\gamma$-ray emission has been found from the direction of B1957+20 in the analysis of the MAGIC data (see Fig.~\ref{fig:th2}).
For the zenith angle distribution of the B1957+20 observations, these cuts correspond to an energy threshold of $\sim\,250$ GeV.
There is \kom{also} no significant emission in the region around the nominal position of the source, as seen in the skymap of Fig.~\ref{fig:skymap}.

\begin{figure}
  \includegraphics[width=0.49\textwidth]{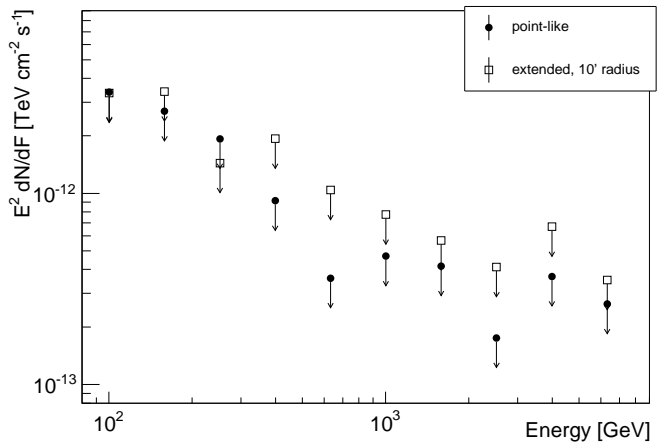}
\caption{
95\% confidence level upper limits on the level of spectral energy distribution of B1957+20 computed under the assumption of a point-like source (filled circles) and an extended source with 10\,arcmin radius (empty squares).}
\label{fig:ulsed}
\end{figure}

Let us first assume that we can treat the source as point-like (i.e. its radius should be much smaller then the angular resolution of MAGIC at those energies, $\sim0.07^\circ$).
Using the whole data set we obtain an upper limit on the flux above 200\,GeV of $3.0 \times 10^{-12}\,\mathrm{cm^{-2}\,s^{-1}}$ (i.e. $\sim1.3\%$\,C.U.). 
This value is rather high comparing with the sensitivity of MAGIC telescopes above those energies due to a small positive excess which raises the value of the upper limit. 
Using the complete data set we also computed upper limits on the flux in
differential energy bins (5 per decade). We present them in the spectral energy
distribution (SED) form in Fig.~\ref{fig:ulsed}.

\begin{figure}
  \includegraphics[width=0.49\textwidth]{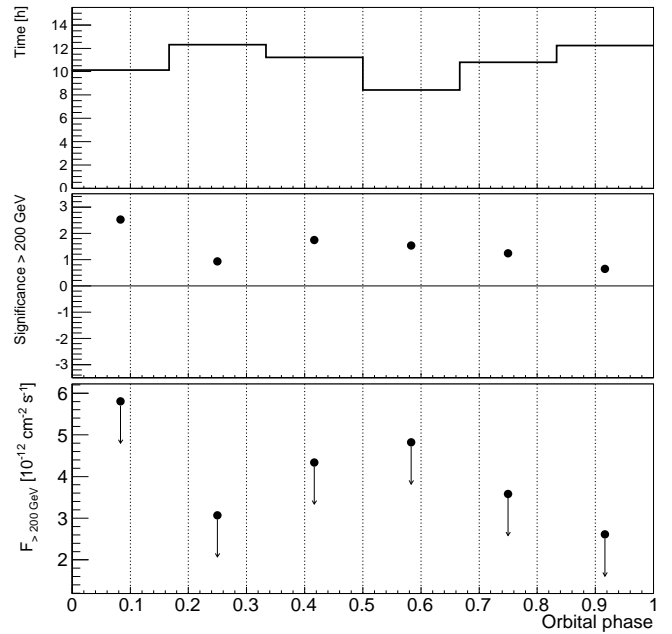}
\caption{
Analysis of B1957+20 in bins of orbital phase: total observation time \kom{spent} at a given phase (top panel), 
the corresponding significance of excess (middle panel) and the 95\% confidence level upper limit on the integral flux (bottom panel) above 200\,GeV. }
\label{fig:ulphase}
\end{figure}

We performed also an analysis dividing the total data set into 6 bins according to the orbital phase of the binary system. As can be seen in top panel of Fig.~\ref{fig:ulphase}, MAGIC observations covered homogeneously the whole orbit of the system. 
The highest obtained significance, $2.5\sigma$, occurred in orbital phase bin of 0-0.17.
After correcting for the number of trials corresponding to the binning in the orbital phase, the significance drops to $1.8\sigma$, making it consistent with a random fluctuation. 
Values of the upper limits in individual bins of orbital phase range from $2.6 \times 10^{-12}\,\mathrm{cm^{-2}\,s^{-1}}$ ($1.1\%$\,C.U.) up to $5.8 \times 10^{-12}\,\mathrm{cm^{-2}\,s^{-1}}$ ($2.6\%$\,C.U.).

According to some of the models the emission from the Black Widow nebula might appear as slightly extended for MAGIC.
Assuming a 10 arcmin radius of the source with a flat-top distribution we have computed the significance of the excess within a broader $\theta^2$ cut and obtained a slightly higher value of $2.2 \sigma$.
As more background events are integrated with a broader cut, the upper limit on the flux from the extended source is worse,  $3.7 \times 10^{-12}\,\mathrm{cm^{-2}\,s^{-1}}$ (i.e. $\sim1.6\%$\,C.U.)  above 200\,GeV. 
We have also calculated the upper limit of the flux above 200\,GeV of such an extended source shifted by $\sim$13 arcmin (see Eq.~\ref{eqD}) opposite to the direction of the observed movement of the binary and obtained $1.8 \times 10^{-12}\,\mathrm{cm^{-2}\,s^{-1}}$.

\subsection{Analysis of the pulsed emission}
\begin{figure}
  \includegraphics[width=0.49\textwidth, trim=0 38 0 35, clip]{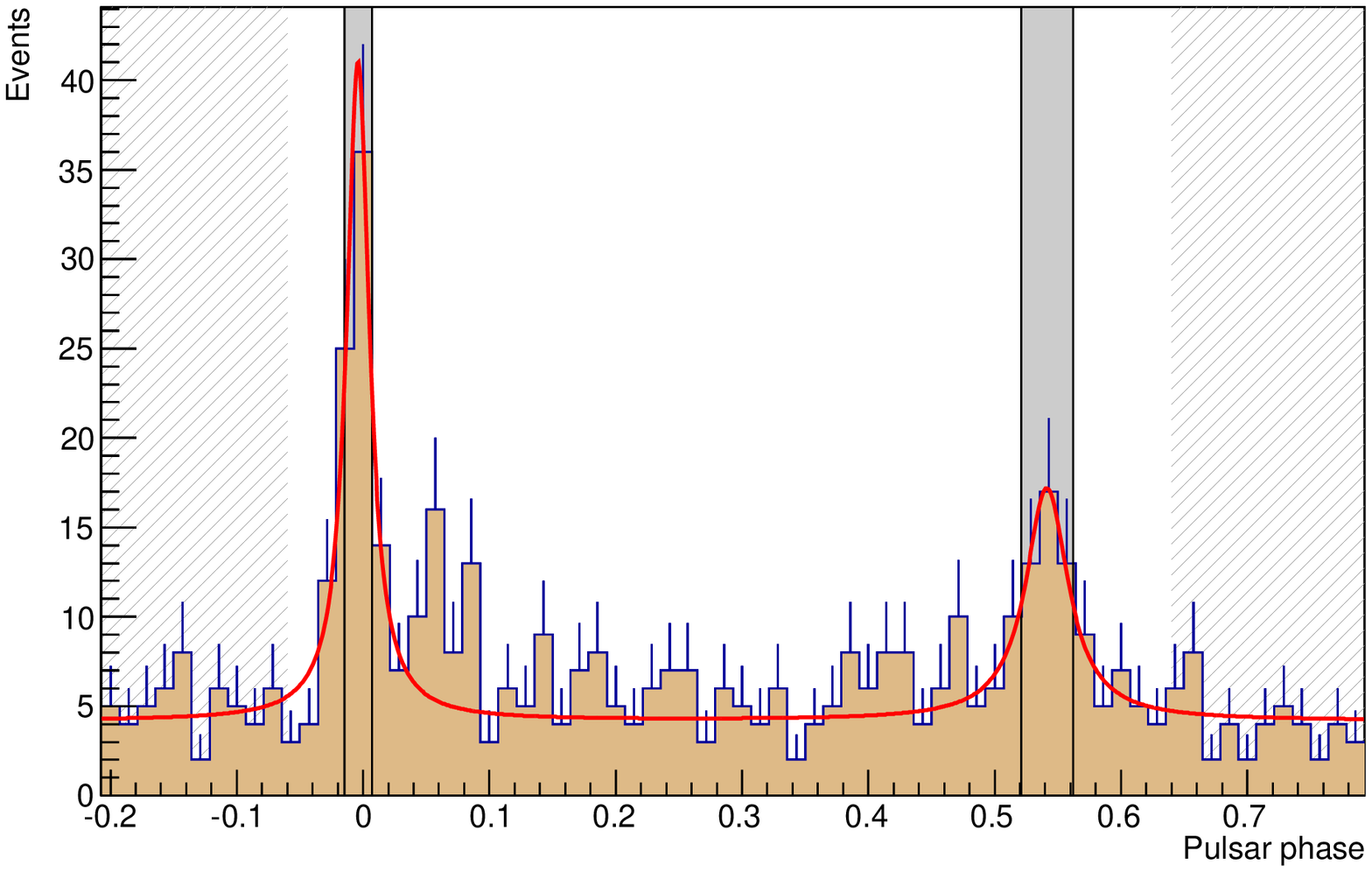}\\
  \includegraphics[width=0.49\textwidth, trim=0 0 0 35, clip]{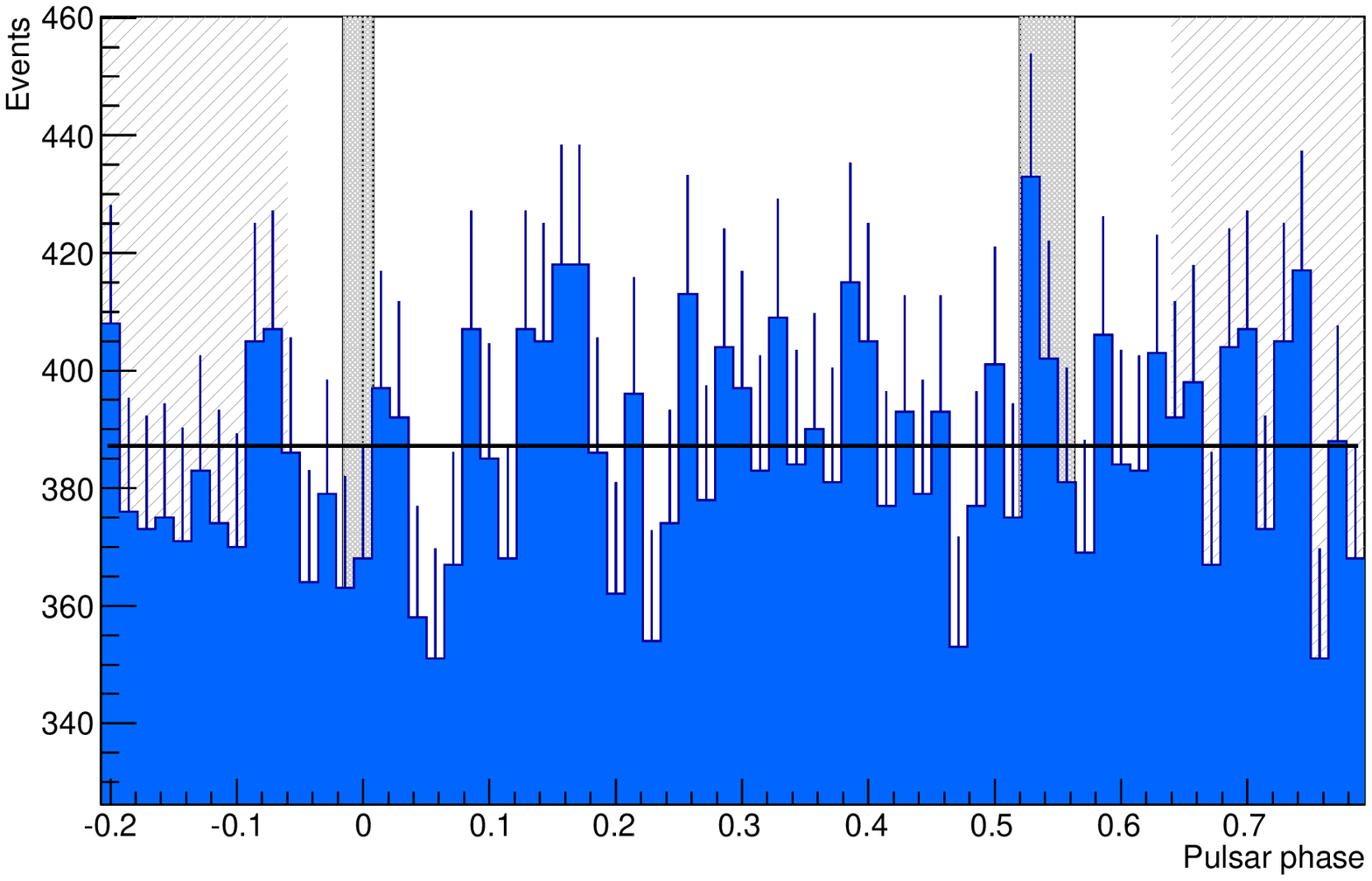}
\caption{ 
Light curve of the $\gamma$-ray emission from PSR B1957+20 using \textit{Fermi}-LAT data above 1 GeV (top panel) and 
MAGIC above 50 GeV using only data with zenith angles $<30^\circ$ (bottom panel). 
The phase 0 corresponds to the position the main peak seen at radio wavelengths. 
The red line in the top panel represents the fitted Lorentzian functions. 
The shaded areas correspond to the phase intervals within the full widths at half-maximum of the fitted peaks, while the dashed regions represent the regions used to estimate the unpulsed background level. 
The horizontal black line in the bottom panel represents the fitted background level.
}
\label{fig:FermiMAGICLC}
\end{figure}

To determine the characteristics of the B1957+20 light curve at energies as close as possible to the MAGIC energy range, we have phase folded the {\it Fermi}-LAT photons with energies above 1 GeV.  
The resulting light curve, after applying the energy dependent angular cut described in Sect.~\ref{sec:fermiana}, is shown in
top panel of Fig. \ref{fig:FermiMAGICLC}. Above this energy the two peaks are still clearly visible.
We obtained a value of the bin-independent {\it H-test} parameter \citep{deJagerHtest} of 179.5 corresponding to a  pulsation significance of $11.8\sigma$. 
Each peak was fitted with a Lorentzian symmetric function. 
The signal regions for the analysis of the MAGIC data
 were defined as the regions within the full widths at half-maximum of the
 fitted peaks. 
We considered as background regions the off-pulse, where no emission is expected
from the pulsar, starting 2.5 FWHM away from the peak centers. 
Table \ref{peakRegions} summarizes the signal and background regions later used for the analysis of
MAGIC data. From now on P1 and P2 will always be referred to as the values in Table \ref{peakRegions}.

\begin{figure}
  \includegraphics[width=0.49\textwidth]{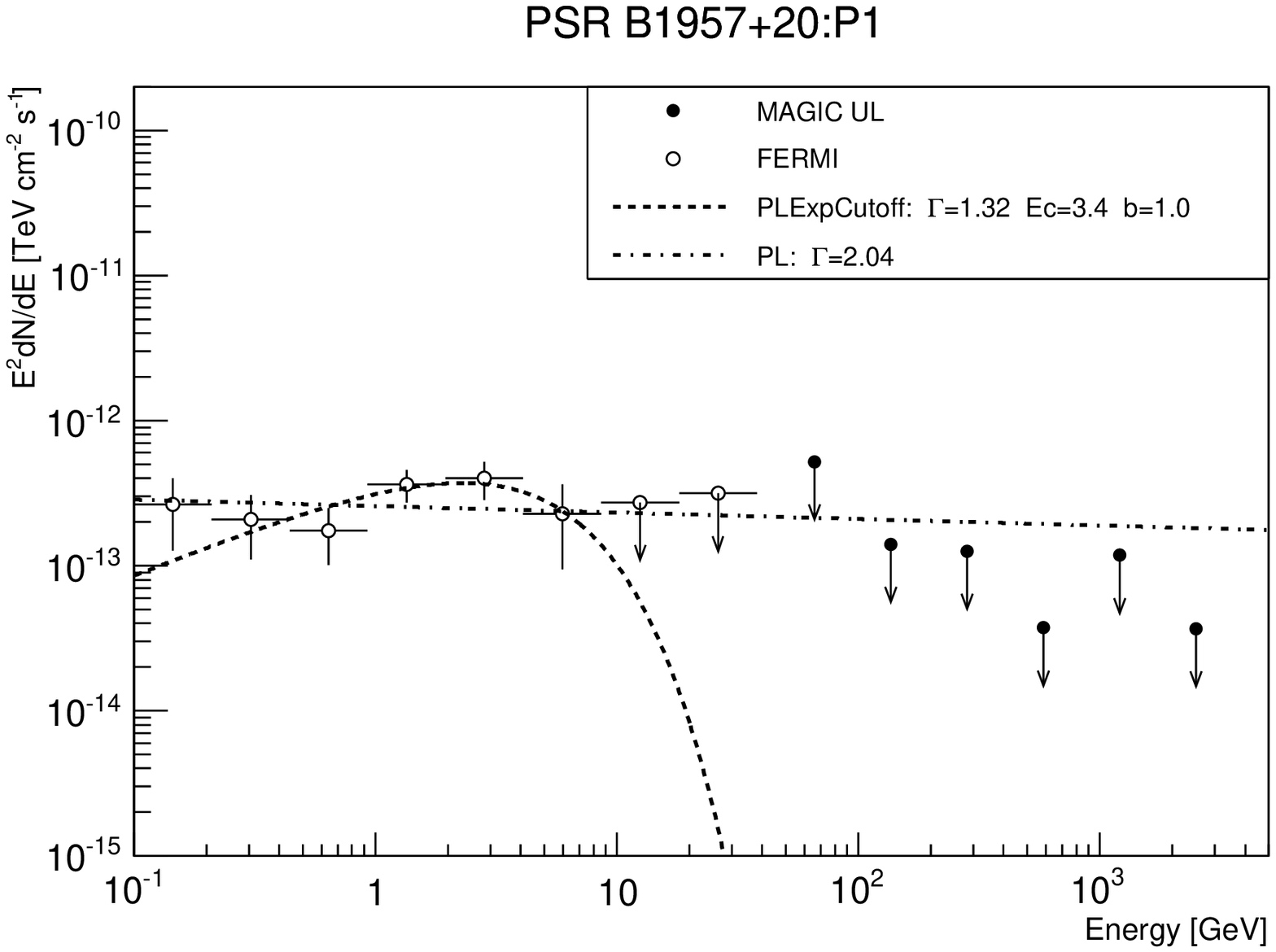}\\
  \includegraphics[width=0.49\textwidth]{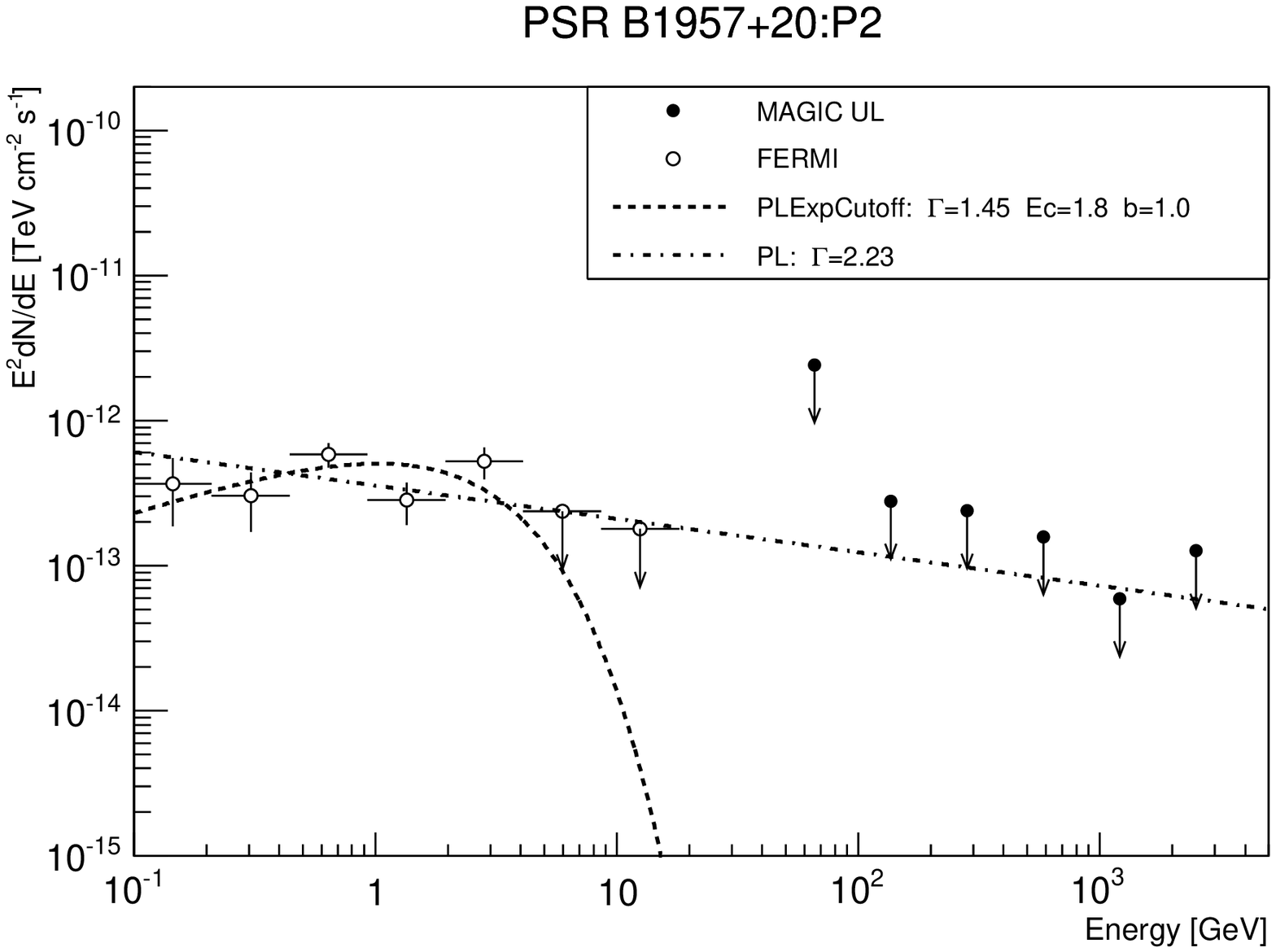}
\caption{MAGIC upper limits for pulsed emission on the level of spectral energy
  distribution of B1957+20 and \textit{Fermi}-LAT phase spectral energy points 
  for P1 (top panel) and P2 (lower panel). }
\label{fig:ulpulsar}
\end{figure}

We fit the {\it Fermi}-LAT spectra above 100 MeV for P1 and P2 
 using a power-law  with a  cut-off function, defined as:
\begin{equation}\label{eq:ecutoff}
  \frac{dF}{dE} =
  N_{0}\left(\frac{E}{E_{0}}\right)^{-\alpha}\exp(-(E/E_{c})^{b}),
\end{equation}
where $E_{0}=0.8$\,GeV is the energy scale, 
$\alpha$ the spectral index, $E_{c}$ the cut-off energy and $b$ the cut-off strength. 
In order to characterize the emission at high energies, we have considered two 
different cases: a simple power-law ($b=0$) and  
a power-law
with an exponential cut-off ($b=1$) \footnote{We tried as well to fit the
    spectra leaving the $b$ parameter of Eq.~\ref{eq:ecutoff} free, but the sharp
    cut-off did not allow us to constrain $b$.}. 
The results of the computed spectra in the phase regions defined in Table~\ref{peakRegions} are shown in Fig.~\ref{fig:ulpulsar} and tabulated in Table~\ref{table:fermiSpectra}.

The MAGIC light curve for the pulsed emission above 50 GeV is shown in the bottom panel of Fig. \ref{fig:FermiMAGICLC}.
No significant pulsation was found in the MAGIC data. 
The total number of excess events in the P1 and P2 phase regions are $23 \pm 47$, 
which corresponds to a significance of $0.5\sigma$.
The significances for P1, P2, and for the sum of both peaks  are tabulated in Table \ref{peakSig} for different energy
ranges.

\begin{table}
 \caption{\small Spectral parameters of the fits of the {\it Fermi}-LAT data above 100 MeV.}
 \begin{center}
\begin{tabular}{ccccc}
 \hline 
 & $N_{0}$  & $\alpha$ & $E_{c}$ (GeV)  & $b$ \\
\hline
 P1 & $4.1 \pm 0.6$ & $2.04 \pm 0.11$ & - & 0\\
 P2 & $5.9 \pm 0.7$  & $2.23 \pm 0.10$ & - & 0\\
 \hline
 P1 & $ 5.6 \pm 1.3$ &  $1.32 \pm 0.35$ &  $3.4 \pm 1.8$  & 1\\
 P2 & $ 12.0  \pm 3.2$ &  $1.45 \pm 0.31$ & $1.8 \pm 0.7$ & 1\\
 \hline
\end{tabular}
\end{center}
{\footnotesize Note: The normalization factor $N_0$ is given in units of $10^{-13}$MeV$^{-1}$s$^{-1}$cm$^{-2}$. The quoted errors are statistical
at a $1\sigma$ confidence level. The systematic errors, as reported by the {\it
  Fermi}-LAT team are of $14\%$ on $\alpha$ and $4\%$ on $E_c$ \citep{2pulsarCatalog}.}
\label{table:fermiSpectra}
 \end{table}

\begin{table}
 \centering 
 \caption{\small Significances of the search of pulsation in the MAGIC data for
   P1, P2 and for the sum of both peaks. 
}
\begin{tabular}{crrr}
 \hline 
Energy range (GeV) & P1+P2 & P1& P2 \\
\hline
$\geq 50$ & $0.49\sigma$ & $-0.91\sigma$ & $1.28\sigma$  \\
$ 50 - 100 $ & $0.45\sigma$ & $-1.36\sigma$ & $1.54\sigma$  \\
$ 100 - 200 $ & $ -0.06\sigma$ & $0.09\sigma$ & $-0.14\sigma$  \\
 \hline
 \end{tabular}
\label{peakSig}
 \end{table}

The MAGIC differential upper limits computed for the pulsed emission are shown
in Fig.~\ref{fig:ulpulsar} by  black arrows.
The spectral indices used for the upper limits computation were obtained from the
power-law fits of the \emph{Fermi}-LAT data shown in Table \ref{table:fermiSpectra}.

\section{Comparison with radiation models}
\kom{TeV $\gamma$-ray emission is already observed from several nebulae around classical pulsars which are formed in explosions of massive stars. 
Such nebulae are usually confined to a relatively small volume by the surrounding dense matter from the supernova remnant or the pre-supernova environment. 
Therefore, the interaction of relativistic electrons with the dense radiation field of soft photons takes place in a small volume around the pulsar. 
On the other hand, the only variable, rotation powered pulsar observed in the TeV $\gamma$-rays (PSR 1259-63, \citealp{aha05}) forms a binary system with a massive Be type star. 
This star creates a strong target for electrons at the periastron passage. 
In the case of binary systems containing MSPs the situation seems to be different. 
As MSPs are only found within low mass binaries, the radiation fields created by their companion star are of lower density.
Consequently, the production of $\gamma$ rays via comptonization process may be less efficient. 
In order to constrain the possible high energy radiation processes around MSPs we performed observations of the classical MSP binary system B1957+20. 
In fact,} observation of non-thermal X-ray emission from \kom{this binary system} indicates a presence of relativistic electrons within the binary system itself and in the extended region around it in the nebula. 
Therefore, some scenarios for the TeV $\gamma$-ray emission from the MSP binary systems were suggested. Moreover, recent discovery of pulsed $\gamma$-ray emission observed up to $\sim$1 TeV from the Crab pulsar \citep{ali08, ali11, ale12, ahn15} stimulates searches for such \kom{pulsed} emission component also in the MSPs. The constraints provided by the observations reported in this paper on $\gamma$-ray emission from B1957+20 are discussed in this section.

\subsection{Gamma-rays from the bow shock nebula}\label{sec:bowshock}

The non-thermal X-ray radiation and H$\alpha$ emission argue that the MSP binary system B1957+20 is surrounded by the bow shock nebula within which relativistic electrons are likely to be accelerated. 
In fact, the observed non-thermal X-ray emission extending up to $\sim$10 keV, if produced in the synchrotron process in the magnetic field of the nebula of the order of a few $\mu$G (expected within the MSP wind nebula), requires the existence of electrons with energies of the order of $\sim$100 TeV \citep{che06,bs13a}. 
The problem appears whether such electrons are able to interact efficiently in the vicinity of the binary system, producing TeV $\gamma$-ray fluxes observable by the present Cherenkov instruments. 
\cite{bs13a} consider the model for such bow shock nebula around the binary system B1957+20 in which electrons are assumed to be accelerated in the pulsar wind region. Electrons diffuse within the large scale nebula (of the order of parsecs) comptonizing the Cosmic Microwave Background and the infrared and optical diffuse radiation in the galactic disk. 
The synchrotron and IC $\gamma$-ray spectra, calculated in terms of this model,
depend on the pulsar and interstellar medium parameters. 
Some of these parameters can be constrained using observations of the non-thermal X-ray emission of this binary system \citep{hua12}.

\begin{figure}
\vskip 6.truecm
\includegraphics{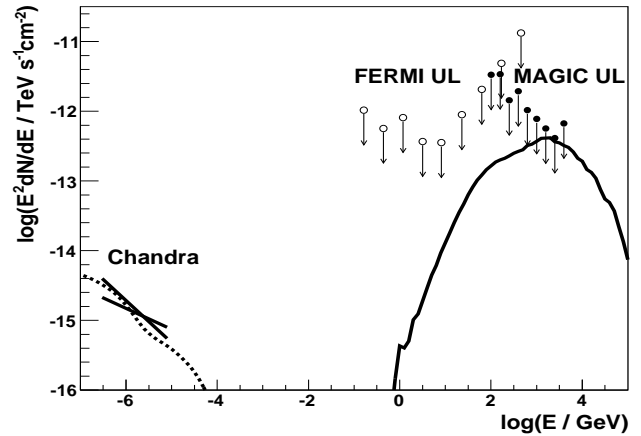}
\caption{Comparison of the upper limits on the $\gamma$-ray emission from the nebula around the Black Widow binary pulsar B1957+20 obtained with MAGIC (10\,arcmin radius extension assumed, filled points), \textit{Fermi}-LAT (point-like, empty points) and the X-ray tail emission detected by {\it Chandra} (\protect\citealp{hua12}, thick bow-tie) with the predictions of the bow shock nebula model (region I) by \protect\cite{bs13a} (solid and dotted lines). 
Assumed model parameters: leptons are injected from the pulsar wind region with a power-law spectrum with an index of 2.5 between 3 and 90 TeV, the magnetization parameter of the pulsar wind  $\sigma = 0.01$, the shock in the wind is located at a distance of $10^{16}$ cm from the pulsar and the minimum magnetic field strength within the nebula is $0.5\mu$G.
}
\label{fig5}
\end{figure}

We first compare the upper limits on the extended TeV $\gamma$-ray emission from the binary system B1957+20 derived in Sect.~\ref{sec:res}, with the $\gamma$-ray spectra expected in terms of the model discussed by \citet{bs13a} under the assumption that the whole X-ray emission is produced within this model. 
\textit{Fermi}-LAT measurement does not constrain the model, as the obtained flux upper limits overshoot model curve by an order of magnitude even for point-like emission.
The model predicts the existence of an extended TeV source towards this binary system with the radius of the order of $\sim$10 pc. 
Assuming the distance to the binary system being 2.5 kpc (consistent with the
lower limit $\sim$2 kpc, \citealp{ker11}), this corresponds to the angular extend
of $\sim$13.8 arcmin.
Moreover, the source is likely to be displaced from the present location of the binary system by the distance determined by the cooling time scale of electrons on the IC scattering of the infrared radiation from the galactic disk and the fast movement of the binary system through the interstellar space
(see Sect.~2 in \citet{bs13a}). 

The dimension of this $\gamma$-ray source is then estimated on 
$D\approx 8.5\times 10^7/\gamma_{\rm e}\approx 42.5 (E_{\rm e}/{\rm TeV})^{-1}$~pc. 
It can be expressed 
by the energy of produced $\gamma$ rays applying the approximate relation between the energies of electrons and $\gamma$ rays produced by them, 
$E_\gamma/{\rm TeV} = \varepsilon\gamma_{\rm e}^2\approx 0.04(E_{\rm e}/{\rm TeV})^{2}$, 
assuming an energy of the infrared photons equal to $\varepsilon\approx 0.01$ eV.
Then, the diameter of the source is estimated on: 
\begin{equation}
D\approx 8.5 /\sqrt{E_\gamma/{\rm TeV}}\,\mathrm{pc}. \label{eqD}
\end{equation}
Such $\gamma$-ray source at the distance of 2.5 kpc has an angular radius of $\sim$6 arc min
at energy of 1 TeV, and thus is well contained within the assumed 10 arc min radius of the extended source analysis.

The synchrotron X-ray spectrum from the tail in the nebula, extending up to $\sim10$ keV, puts constraints on the magnetization parameter of the pulsar wind, $\sigma\gtrsim0.01$ (see Sect.~3 in \citealp{bs13a}). 
In this model, the synchrotron emission and inverse Compton (IC) $\gamma$-ray emission from the extended nebula is produced by the same leptons. Therefore, the observed  non-thermal X-ray emission can be used as the normalization of the electron content within the nebula. We estimate that about $10\%$ of the pulsar spin down energy has to be converted to the relativistic leptons in the nebula. 
Still, the \kom{TeV emission predicted in this model lies} below the MAGIC upper limits, derived for the extended source with a radius of 10 arcmin (see Fig.~\ref{fig5}).
The predictions of the model are barely consistent with the obtained upper limits on the flux.

\subsection{Gamma-rays from the inner mixed wind nebula}\label{sec:innernebula}

In general, nebulae around black widow MSP binary systems are expected to have quite complicated structure since, due to the Coriolis force, the pulsar wind can partially mix with the wind from the companion star in the equatorial region of the binary system (e.g. \citealp{bs13b, zab13}). 
Therefore, it is expected that in this region, the pulsar wind overtakes the slow stellar wind creating the collision region at some distance from the binary system. Then, the pulsar wind becomes loaded with the matter of the stellar wind. 
The mixed winds move together relatively slowly away from the binary system. 
Therefore, electrons, accelerated or captured in such a collision region of the winds close to the radiation field of the companion star, can comptonize efficiently the radiation from the hot part of the companion star. 
As a result, more efficient production of the high energy $\gamma$ rays is expected than in the case of rectilinear propagation of leptons through the binary system (see Table~2 in \citealp{bs13b} for parameters describing such scenario in the case of B1957+20). Since the collision region is outside of the binary system, this TeV $\gamma$-ray emission should be steady, point-like (i.e at a distance of 
$\sim 2\times10^{12}$\,cm or larger from the binary system) and centered on the present location of the binary system B1957+20. 

\begin{figure}
\vskip 6.truecm
\includegraphics{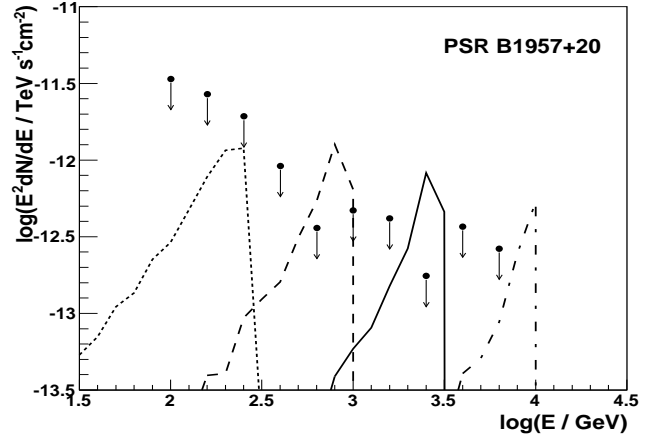}
\caption{The comparison of the upper limits on the point-like TeV $\gamma$-ray emission from the inner part of nebula around the Black Widow binary pulsar B1957+20 with the predictions of the mixed wind nebula model (region II). 
The leptons in the model are accelerated with the mono-energetic spectrum with energies equal to 300 GeV (dotted curve), 1 TeV (dashed), 3 TeV (solid), and 10 TeV (dot-dashed). It is assumed that the factor, describing the level of $\gamma$-ray emission, is equal to $\Delta_{\rm mix}\varepsilon = 0.1$ (see Sect.~4.2 for details). 
}
\label{fig6}
\end{figure}

Following the model of \citealp{bs13b} we performed numerical calculations of expected TeV $\gamma$-ray fluxes assuming that electrons are injected into the mixed wind region from the pulsar (or its surroundings) with a close to monoenergetic spectrum. Leptons with the TeV energies are predicted to escape from the inner pulsar magnetospheres (e.g. \citealp{stu71,rs75}). 
The electron spectra, dominated energetically at the highest energies, are also expected to originate in the close vicinity of the pulsars. For example, the GeV $\gamma$-ray synchrotron flares from the Crab nebula show very flat spectra which have to be produced by leptons with spectra dominated at the highest energies (\citealp{tav11,ab11}).
The level of $\gamma$-ray emission in this model is determined by the product, $\Delta_{\rm mix}\varepsilon$, of the solid angle covered by the mixed winds, $\Delta_{\rm mix}$, and the energy conversion efficiency from the pulsar to relativistic electrons, $\varepsilon$. 
In the case of B1957+20, $\Delta_{\rm mix}$ is estimated on $\sim$0.82 (\citealp{bs13b}). We compare the upper limits derived for the case of the point-like TeV $\gamma$-ray source towards B1957+20 with the results of calculations of the TeV $\gamma$-ray fluxes for $\Delta_{\rm mix}\varepsilon = 0.1$ (see Fig.~\ref{fig6}) and other basic parameters of B1957+20 following the Table 1 and 2 in \citet{bs13b}.
Based on this comparison we estimate the upper limit on the acceleration efficiency of electrons in terms of such a model in the range $\varepsilon\leq 2-10\%$ for the mono-energetic electrons with energies in the range 0.3-10 TeV.
Thus, the acceleration process of mono-energetic leptons to energies in the range covered by the MAGIC telescope energy range cannot be very efficient. 
We have also calculated the TeV $\gamma$-ray fluxes for B1957+20 in the case of injection of electrons with the power-law spectrum. 
They are clearly on a lower level than the upper limits on the TeV $\gamma$-ray point-like source towards B1957+20 (in Fig.~\ref{fig:ulsed}). 
Therefore, the point-like TeV 
$\gamma$-ray emission from B1957+20, produced by electrons with the power-law spectrum in terms of this model, is predicted to be below the upper limits derived here based on the MAGIC telescopes observations.

\subsection{Modulated gamma-rays from the binary system}\label{sec:innerbinary}

The MSP binary systems of the black widow type are compact enough that the energy released by the pulsar can be responsible for the effective evaporation of the companion star and production of a relatively strong stellar wind (e.g. \citealp{rud89}).
Moreover, the pulsar efficiently heats up a part of the stellar surface.
The increased temperature in part of the stellar surface results in a modulation of the optical emission from the companion star (e.g. \citealp{bre13}).
The interaction of the pulsar and stellar winds creates a transition region (a shock structure) at a distance $\sim$5 times larger than the size of the companion star ($10^{10}$\,cm) in which particles can be accelerated \citep{hg90, at93, whvb15}. It has been recently argued that the stellar and pulsar winds can mix efficiently in this transition region. Then, accelerated electrons might have enough time to comptonize stellar radiation to the TeV energies and produce detectable $\gamma$-ray fluxes \citep{bed14}.  In such case, it is expected that the radiation processes within MSP binary systems could occur similarly to those observed in the TeV $\gamma$-ray binaries such as LS5039 or LSI 303 +61.
The maximum energy of electrons accelerated in such wind collision region is limited by their advection from the wind collision region and/or the synchrotron process. 
They depend on a few free parameters of the model from which two, namely the magnetization parameter of the pulsar wind, $\sigma$, and the velocity of the mixed pulsar/stellar wind,  are not presently constrained by the observations. The $\sigma$ parameter can not be too large in order not to dominate the electron energy losses by the synchrotron process. On the other hand, it cannot be also too low  in order to allow acceleration of electrons to TeV energies.

\begin{figure}
\vskip 6.truecm
\includegraphics{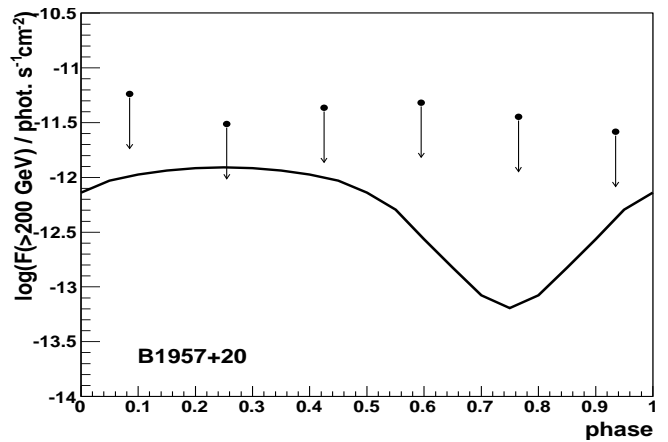}
\caption{The comparison of the $\gamma$-ray light curve, folded with the orbital period of the system, at energies above 200 GeV, expected from the MSP binary system B1957+20 in terms of the IC model (region III) discussed by \citet{bed14}, with the upper limits on the $\gamma$-ray flux from this binary system at specific range of phases. The predicted $\gamma$-ray fluxes are obtained from the normalization of the synchrotron spectra, produced by the same population of electrons, to the observed modulated X-ray emission from this binary (see \citealp{hua12}). 
}
\label{fig7}
\end{figure}

In order to interpret the MAGIC measurement of B1957+20 we choose a magnetization parameter $\sigma = 10^{-4}$, a mixed wind velocity $v_{\rm w} = 10^{10}$ cm s$^{-1}$, and other parameters of the model following 
\citep{bed14}. The model assumes that electrons in the wind collision region are accelerated with a power-law spectrum with spectral index $-2$ and that $1\%$ of the rotational energy loss rate is transferred to relativistic electrons.
Note that the IC $\gamma$-ray emission expected in this model does not depend strongly on the values of $\sigma$ and $v_{\rm w}$ in contrast to the synchrotron emission (see Fig.~2ab in \citealp{bed14}). 
The $\gamma$-ray emission produced by these electrons is expected to be modulated with the period of the binary system since the stellar radiation field is anisotropic with respect to the acceleration region of electrons. 
It was predicted to be on the sensitivity limit of the extensive observations with the current generation of the Cherenkov telescopes at the binary phase when the pulsar is behind the companion star \citep{bed14}. We have performed calculations of the $\gamma$-ray emission in terms of this model for the MSP binary system B1957+20 and for the set of the parameters described above see also Table 1 in \cite{bed14}. 
In Fig.~\ref{fig7} we compare the predicted, orbital phase folded, $\gamma$-ray light curve with the upper limits obtained by MAGIC. The largest TeV $\gamma$-ray fluxes are predicted by the model for the orbital phase of 0.25 (i.e. when the pulsar is behind the star).  The integral $\gamma$-ray flux above 200 GeV expected in this model, equal to $1.4\times 10^{-12}$ cm$^{-2}$ s$^{-1}$ is below a MAGIC upper limit of $3.1\times 10^{-12}$ cm$^{-2}$ s$^{-1}$. Therefore, the present observations with the MAGIC telescopes allow us to conclude that less than $3\%$ of the rotational energy loss rate of the pulsar within the binary system B1957+20 has been transferred to relativistic electrons accelerated at the region of collision of the pulsar and stellar winds as postulated in above mentioned model.

\subsection{Pulsed gamma-rays from the pulsar}\label{sec:pulsar}

Recent discovery of pulsed $\sim$100 GeV $\gamma$-ray emission from the Crab pulsar \citep{ali11, ale12, ahn15} provided new constraints on the models for the GeV-TeV $\gamma$-ray emission from the pulsar magnetospheres. Different scenarios and radiation mechanisms have been proposed as responsible for this unexpected emission. 
For example, \citet{lom12} and \citet{ale12} argue that this pulsed emission component is produced in the comptonization process occurring within the outer gap scenario. 
It is also suggested that electrons, accelerated close to the light cylinder radius (at a distance of $7\times 10^6$\,cm from the Black Widow pulsar), can produce $\gamma$ rays in curvature process in the distorted magnetic field (\citealp{bed12, bog14}) or due to IC scattering of thermal radiation from the NS surface (\citealp{or17}).    
This emission could also come from the outside of the light cylinder radius, from the region of the equatorial current sheet (e.g. \citealp{ad13, us14, cb14}), or at the distance of a few tens of the light cylinder radii as a result of comptonization of the X-rays produced in the inner magnetosphere \citep{abk12, pet12}. 
Moreover, \cite{bed12} suggests that the middle aged pulsars of the Vela type and some extreme millisecond pulsars should be also able to produce tails of pulsed high energy $\gamma$-ray emission extending to $\sim50$\,GeV.
In fact such emission has been recently discovered from the Vela pulsar \citep{ga15}.
However, such tails are not expected in the case of the older pulsars (e.g. Geminga type). 

In Fig. \ref{fig:ulpulsar} we compare the TeV upper limits obtained by MAGIC on the pulsed $\gamma$-ray emission from B1957+20 with the SED measured by \textit{Fermi}-LAT. 
The MAGIC upper limits are well above the extrapolation of the fitted exponential cut-offs for any of the two emission peaks.

\section{Conclusions}

No signal at the TeV energies has been detected from the Black Widow binary system B1957+20 and its surrounding by the MAGIC telescopes. 
The upper limits above 200 GeV obtained from these data are $\sim$1.3$\%$ C.U. for a point-like source and $\sim$1.6$\%$ C.U. assuming its extension of 10 arcmin. 
They are not very restrictive due to a small positive excess in the direction of B1957+20. 
The orbital phase analysis also does not show any clear hint of emission localized at any particular orbital phase.
The pulsar phase analysis also does not show any positive excess above 50 GeV. 
The derived upper limits for pulsed emission do not allow us to constrain a potential very high energy extension of the \textit{Fermi}-LAT spectra.
We have compared these upper limits with the predictions of a few models which postulate production of
the TeV $\gamma$ rays within different parts of the binary system B1957+20
or its vicinity. From the comparison of the model for the extended bow shock nebula around the Black Widow binary system (see Sect.~5.1), we confirm that the magnetization parameter of the pulsar wind in such an extended nebula formed by the freely expanding wind should be larger than 
$\sigma\sim0.01$. Thus, the magnetization of the winds around MSPs seems to be clearly larger than  
the magnetization of the winds produced by young classical pulsars of the Crab type, equal to $\sigma\sim$0.003 \citep{kc84}, but more similar to the magnetization of the Vela type pulsars in the range $0.05 < \sigma < 0.5$ \citep{sd03}. 
On the other hand, the transfer of energy from the magnetized wind to relativistic leptons might occur more efficiently in the equatorial region of the binary system where the pulsar and stellar winds can mix efficiently (see Sect.~5.2).  
The MAGIC upper limit, derived on a point-like source in the direction of B1957+20, allows \kom{us} to constrain the efficiency of energy conversion from the pulsar to relativistic electrons in this scenario to be below $\sim(2-10)\%$, provided that they are mono-energetic with energies in the range 0.3--10 TeV. The TeV $\gamma$ rays are expected to be also efficiently produced 
in the region of collisions of the pulsar and companion star winds within the binary system (see Sect.~5.3). The largest 
$\gamma$-ray fluxes are predicted in this scenario when the pulsar is behind the companion star. However, The MAGIC 95\% confidence level upper limits are a factor of $\sim$2 above predictions of such a wind collision model for the case of efficient conversion of the pulsar wind energy into relativistic leptons in the collision region and likely parameters of the companion star wind. Therefore, more sensitive observations of this binary system with the future CTA (further supported by a weak hint of excess seen by MAGIC at the level of $\sim2\sigma$) will be needed in order to detect this modulated $\gamma$-ray emission from B1957+20. 
In the case of $\gamma$-ray emission from pulsar itself, MAGIC observations rule out a simple power-law extrapolation of the P1 pulsed emission through the sub-TeV energy range (Sect.~5.4). 
We conclude that MAGIC upper limits on the $\gamma$-ray emission from the MSP binary B1957+20 are generally consistent with the predictions of the present models. Most of these models predict $\gamma$-ray fluxes which depend on the strength of the radiation field created by the companion star. Therefore, future, more sensitive observations of the MSP binary systems of the Redback type (e.g. such as PSR J1816+4510) are expected to provide stronger constraints on the acceleration processes of leptons within the binaries and their surrounding.

\section*{Acknowledgements}
We would like to thank the Instituto de Astrof\'{\i}sica de Canarias for the excellent working conditions at the Observatorio del Roque de los Muchachos in La Palma. The financial support of the German BMBF and MPG, the Italian INFN and INAF, the Swiss National Fund SNF, the ERDF under the Spanish MINECO (FPA2015-69818-P, FPA2012-36668, FPA2015-68\kom{3}78-P, FPA2015-69210-C6-2-R, FPA2015-69210-C6-4-R, FPA2015-69210-C6-6-R, AYA2015-71042-P, \kom{AYA2016-76012-C3-1-P}, ESP2015-71662-C2-2-P, CSD2009-00064), and the Japanese JSPS and MEXT is gratefully acknowledged. This work was also supported by the Spanish Centro de Excelencia ``Severo Ochoa'' SEV-2012-0234 and SEV-2015-0548, and Unidad de Excelencia ``Mar\'{\i}a de Maeztu'' MDM-2014-0369, by the Croatian Science Foundation (HrZZ) Project 09/176 and the University of Rijeka Project 13.12.1.3.02, by the DFG Collaborative Research Centers SFB823/C4 and SFB876/C3, and by the Polish MNiSzW grant \kom{2016/22/M/ST9/00382}.
This work is partially supported by the grant through the Polish Narodowe Centrum Nauki No. 2014/15/B/ST9/04043. 
The Nan\c{c}ay Radio Observatory is operated by the Paris Observatory, associated with the French Centre National de la Recherche Scientifique (CNRS).

%
%


\vspace*{0.5cm}
\noindent
$^{1}$ {ETH Zurich, CH-8093 Zurich, Switzerland} \\
$^{2}$ {Universit\`a di Udine, and INFN Trieste, I-33100 Udine, Italy} \\
$^{3}$ {INAF National Institute for Astrophysics, I-00136 Rome, Italy} \\
$^{4}$ {Universit\`a di Padova and INFN, I-35131 Padova, Italy} \\
$^{5}$ {Croatian MAGIC Consortium, Rudjer Boskovic Institute, University of Rijeka, University of Split - FESB, University of Zagreb - FER, University of Osijek,Croatia} \\
$^{6}$ {Saha Institute of Nuclear Physics, 1/AF Bidhannagar, Salt Lake, Sector-1, Kolkata 700064, India} \\
$^{7}$ {Max-Planck-Institut f\"ur Physik, D-80805 M\"unchen, Germany} \\
$^{8}$ {Universidad Complutense, E-28040 Madrid, Spain} \\
$^{9}$ {Inst. de Astrof\'isica de Canarias, E-38200 La Laguna, Tenerife, Spain} \\
$^{10}$ {Universidad de La Laguna, Dpto. Astrof\'isica, E-38206 La Laguna, Tenerife, Spain} \\
$^{11}$ {University of \L\'od\'z, PL-90236 Lodz, Poland} \\
$^{12}$ {Deutsches Elektronen-Synchrotron (DESY), D-15738 Zeuthen, Germany} \\
$^{13}$ {Institut de Fisica d'Altes Energies (IFAE), The Barcelona Institute of Science and Technology, Campus UAB, 08193 Bellaterra (Barcelona), Spain} \\
$^{14}$ {Universit\`a  di Siena, and INFN Pisa, I-53100 Siena, Italy} \\
$^{15}$ {Universit\"at W\"urzburg, D-97074 W\"urzburg, Germany} \\
$^{16}$ {Institute for Space Sciences (CSIC/IEEC), E-08193 Barcelona, Spain} \\
$^{17}$ {Technische Universit\"at Dortmund, D-44221 Dortmund, Germany} \\
$^{18}$ {Finnish MAGIC Consortium, Tuorla Observatory, University of Turku and Astronomy Division, University of Oulu, Finland} \\
$^{19}$ {Unitat de F\'isica de les Radiacions, Departament de F\'isica, and CERES-IEEC, Universitat Aut\`onoma de Barcelona, E-08193 Bellaterra, Spain} \\
$^{20}$ {Universitat de Barcelona, ICC, IEEC-UB, E-08028 Barcelona, Spain} \\
$^{21}$ {Japanese MAGIC Consortium, ICRR, The University of Tokyo, Department of Physics and Hakubi Center, Kyoto University, Tokai University, The University of Tokushima, Japan} \\
$^{22}$ {Inst. for Nucl. Research and Nucl. Energy, BG-1784 Sofia, Bulgaria} \\
$^{23}$ {Universit\`a di Pisa, and INFN Pisa, I-56126 Pisa, Italy} \\
$^{24}$ {ICREA and Institute for Space Sciences (CSIC/IEEC), E-08193 Barcelona, Spain} \\
$^{25}$ {also at the Department of Physics of Kyoto University, Japan} \\
$^{26}$ {now at Centro Brasileiro de Pesquisas F\'isicas (CBPF/MCTI), R. Dr. Xavier Sigaud, 150 - Urca, Rio de Janeiro - RJ, 22290-180, Brazil} \\
$^{27}$ {now at NASA Goddard Space Flight Center, Greenbelt, MD 20771, USA} \\
$^{28}$ {Department of Physics and Department of Astronomy, University of Maryland, College Park, MD 20742, USA} \\
$^{29}$ {Humboldt University of Berlin, Institut f\"ur Physik Newtonstr. 15, 12489 Berlin Germany} \\
$^{30}$ {also at University of Trieste} \\
$^{31}$ {now at Ecole polytechnique f\'ed\'erale de Lausanne (EPFL), Lausanne, Switzerland} \\
$^{32}$ {also at Japanese MAGIC Consortium} \\
$^{33}$ {now at Finnish Centre for Astronomy with ESO (FINCA), Turku, Finland} \\
$^{34}$ {also at INAF-Trieste and Dept. of Physics \& Astronomy, University of Bologna} \\
$^{35}$ {Laboratoire de Physique et Chimie de l'Environnement et de l'Espace, LPC2E, CNRS-Universit\'{e}  d'Orl\'{e}ans,  F-45071  Orl\'{e}ans, France} \\
$^{36}$ {Station de Radioastronomie de Nan\c{c}ay, Observatoire de Paris, CNRS/INSU, F-18330 Nan\c{c}ay, France} \\

\bsp	
\label{lastpage}
\end{document}